\documentstyle[11pt,psfig]{article}

\def\AU{\,{\rm AU}}

\def\deg{^\circ}
\def\vphm{\vphantom{$B^{B^B}$}}

\makeatletter

\def\thebibliography#1{\section*{References\markboth
 {REFERENCES}{REFERENCES}}\list
  {}{\settowidth\labelwidth{0pt}\leftmargin\labelwidth    
 \advance\leftmargin\labelsep
 \usecounter{enumi}\@bibsetup}
 \def\newblock{\hskip .11em plus .33em minus -.07em}
 \sloppy\clubpenalty4000\widowpenalty4000
 \sfcode`\.=1000\relax}

\def\@bibsetup{\itemindent=-\leftmargin \itemsep=0pt
 \parsep=0pt  
 }

\def\@biblabel#1{\hfill}

\long\def\@makecaption#1#2{%
  \small
  \vskip\abovecaptionskip
  \sbox\@tempboxa{#1: #2}%
  \ifdim \wd\@tempboxa >\hsize
    #1: #2\par
  \else
    \global \@minipagefalse
    \hb@xt@\hsize{\hfil\box\@tempboxa\hfil}%
  \fi
  \normalsize
  \vskip\belowcaptionskip}

\makeatother

\begin{document}

\title{{\bf Resonances and instabilities in symmetric multistep 
 methods}}
\author{Gerald D. Quinlan\thanks{Mailing address: 
1055 Bay St., \#714, Toronto, Canada M5S 3A3}\\
Canadian Institute for Theoretical Astrophysics\\
University of Toronto}
\date{January 12, 1999}
\maketitle
\begin{abstract}
The symmetric multistep methods developed by Quinlan and Tremaine (1990) are
shown to suffer from resonances and instabilities at special stepsizes when
used to integrate nonlinear equations.  This property of symmetric multistep
methods was missed in previous studies that considered only the linear
stability of the methods.  The resonances and instabilities are worse for
high-order methods than for low-order methods, and the number of bad
stepsizes increases with the number frequencies present in the solution.
Symmetric methods are still recommended for some problems, including
long-term integrations of planetary orbits, but the high-order methods must
be used with caution.
\end{abstract}

\section{Introduction}

In many physical problems one has to solve second-order differential equations
of the type
\begin{equation} \label{eq-eqn}
  x''(t)=f(x,t),
\end{equation}
where $x(t)$ is the position at time $t$ and $f(x,t)$ is the force, assumed
to be independent of the velocity.  Such an equation or set of equations
can be solved efficiently by a multistep method
\begin{equation}  \label{eq-genmulti}
  \alpha_k x_{n+k}+\cdots+\alpha_0 x_n=
  h^2\left(\beta_k f_{n+k}+\cdots+\beta_0 f_n\right),
\end{equation}
where $x_n$ is the computed position at time step $n$ and $h$ is the
stepsize.  A popular class of such methods is the the St\"ormer-Cowell
class, for which $\alpha_k=1$, $\alpha_{k-1}=-2$, $\alpha_{k-2}=1$, and
$\alpha_{k-3}= \cdots = \alpha_0=0$, with the St\"ormer method being
explicit ($\beta_k=0$) and the Cowell method implicit ($\beta_k\neq0$).
St\"ormer-Cowell methods have often been used for long-term integrations of
planetary orbits (see Quinlan and Tremaine 1990 and references therein).
But the St\"ormer-Cowell methods suffer from a defect, sometimes called an
orbital instability, when the stepnumber $k$ exceeds 2: if a
St\"ormer-Cowell method with $k>2$ is used to integrate a circular orbit,
the radius does not remain constant, and the orbit spirals either inwards or
outwards (the direction depends on $k$).  This defect was recognized by
Gautschi (1961) and Stiefel and Bettis (1969), who proposed modified
multistep methods for orbital integrations.  Their methods require {\it a
priori} knowledge of the frequency of the solution, however, which is
usually unknown or, at best, known only approximately.

Lambert and Watson (1976) showed that the orbital instability of the
St\"ormer-Cowell methods can be avoided by choosing the coefficients of a
multistep method to be symmetric, so that
\begin{equation} \label{eq-sydef}
  \alpha_i=\alpha_{k-i},\quad \beta_i=\beta_{k-i},\qquad i=0,\ldots,k.
\end{equation}
Lambert and Watson analysed in detail the application of symmetric methods
to the linear test equation
\begin{equation} \label{eq-sho}
  x''(t)=-\omega^2x(t),
\end{equation}
and showed that if $\omega^2h^2$ lies within an interval $(0,H_0^2)$, which
they called the interval of periodicity, the solution is guaranteed to be
periodic.  Quinlan and Tremaine (1990) extended the work of Lambert and
Watson (1976) to derive high-order explicit symmetric methods suitable for
the integration of planetary orbits, and compared these with high-order
St\"ormer methods.  The symmetric methods gave energy errors that did not
grow with time, and position errors that grew only linearly with time,
whereas the St\"ormer methods gave energy errors that grew linearly with
time, and position errors that grew as the time squared.

\begin{figure}[ht]
\centerline{\psfig{figure=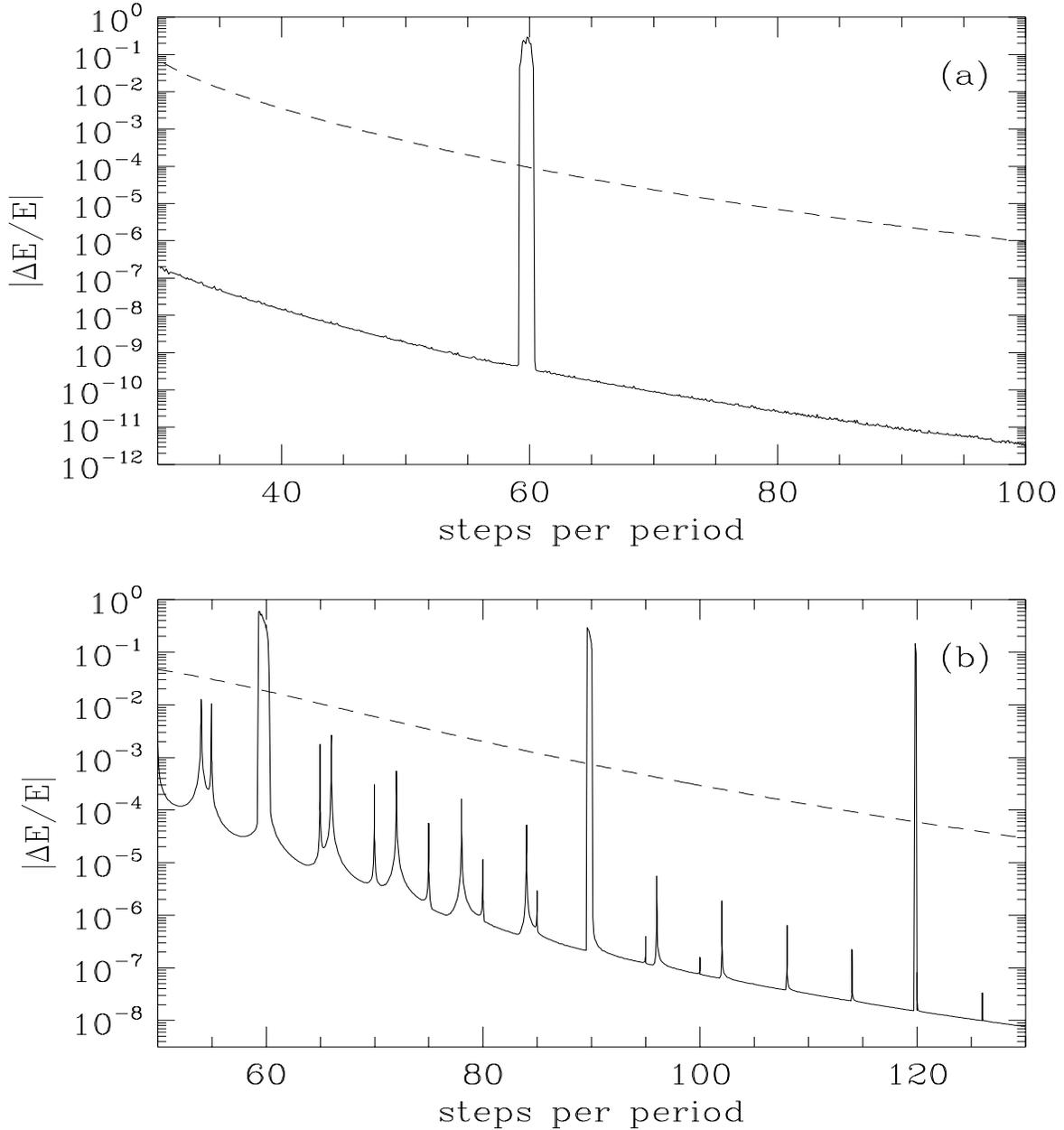,width=\textwidth,height=0.85\textheight}}
\vskip-30pt
\caption[SY8]
{Maximum fractional energy error during integrations of a Kepler orbit for
25000 periods using the method SY8 (solid lines) and the 8th-order St\"ormer
method (dashed lines), plotted as a function of the number of steps per
orbit: (a) circular orbit; (b) eccentric orbit ($e=0.2$).}
\label{fig-sy8}
\end{figure}

\begin{figure}[ht]
\centerline{\psfig{figure=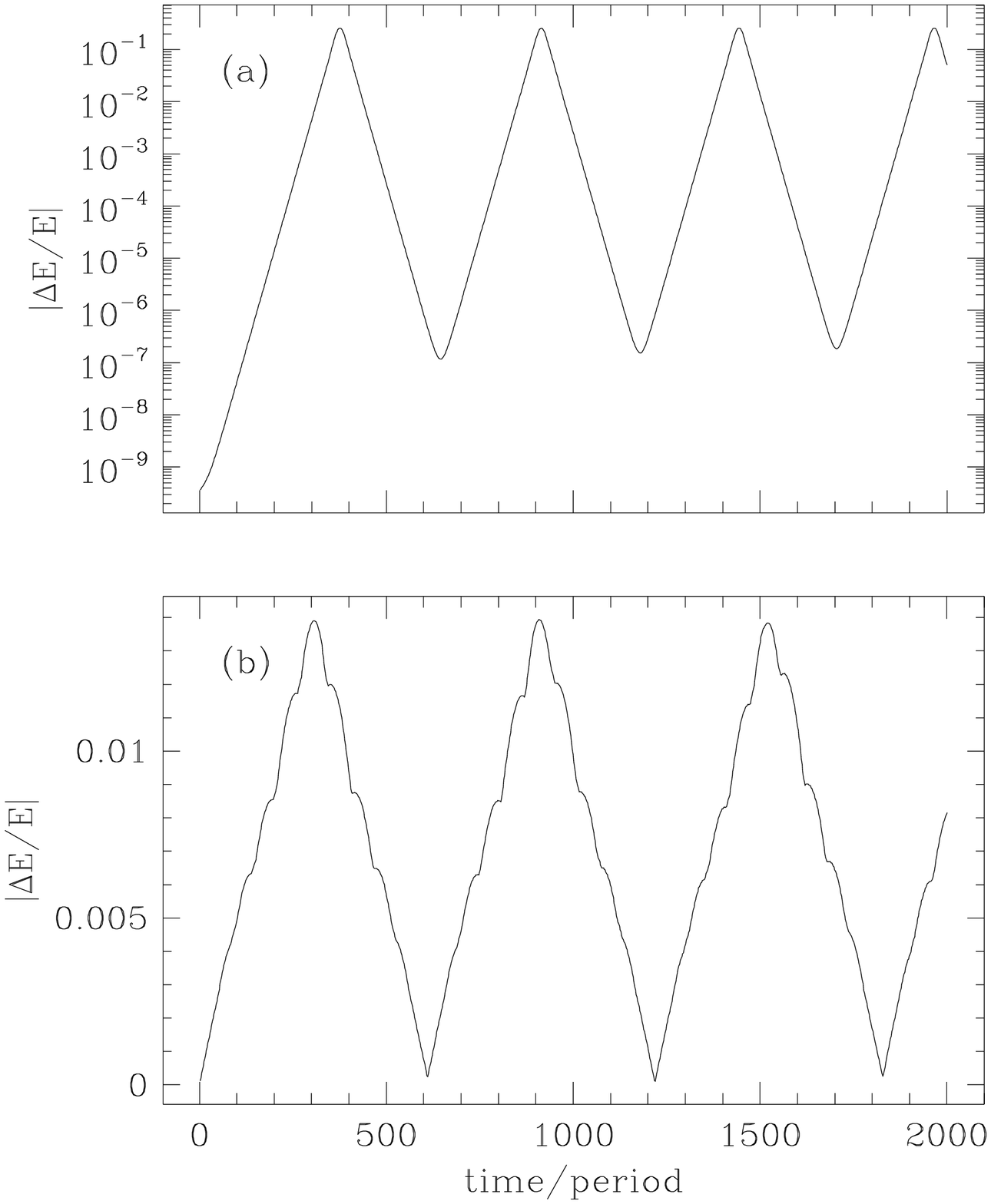,width=\textwidth,height=0.85\textheight}}
\vskip-50pt
\caption[SY8 Time dependence.]{
Time dependence of the fractional energy error with the 8th-order symmetric
method SY8. Panels (a) and (b) plot the maximum error observed in the previous
period during integrations of Kepler orbits with (a) a circular orbit using
60 steps per period and (b) an eccentric orbit ($e=0.2$) using 54.03 steps
per period.} 
\label{fig-growth}
\end{figure}

Soon after Quinlan and Tremaine's methods were published, however, Alar
Toomre discovered a disturbing feature of the methods, an example of which
is shown in Figure~\ref{fig-sy8}.
Panel~(a) shows the maximum error in the energy of a circular Kepler orbit
integrated with the 8th-order symmetric method of Quinlan and Tremaine and
with the 8th-order St\"ormer method, plotted versus the stepsize used in the
integration.  The energy error decreases with the stepsize as $\sim h^9$, as
expected for an 8th-order method, and at most stepsizes the error from the
symmetric method is much smaller than the error from the St\"ormer method.
But there is a startling spike in the energy error from the symmetric method
near a stepsize corresponding to 60 steps per orbit, a stepsize that is well
within this method's interval of periodicity. Figure~\ref{fig-growth} shows
the time development of the instability for a circular orbit integrated with
60 steps per orbit.
The energy error grows exponentially for about the first 400 periods until
it reaches a maximum value\footnote{The value of the energy error at the
instability depends on the formula used to compute the velocities. If the
multistep equation is written in summed form (Henrici 1962; Quinlan 1994)
and the velocities are computed from the summed accelerations, the maximum
error is about 10 times smaller than shown in Figure~\ref{fig-growth}, but
apart from this difference the instability remains the same.  In this paper
the summed form was used only for the planetary integrations described in
Section~\ref{sec-planets}.} of about 0.25. It decreases over the next few
hundred periods to a minimum value of about $10^{-7}$, and then oscillates
between these extremes with a period of roughly 550 orbital periods.  The
longitude error (not shown in the figure) grows exponentially until it
reaches a value of order unity and then stays at that level.

The problems are worse for eccentric orbits, as shown in panel~(b) of
Figure~\ref{fig-sy8} for an orbit with $e=0.2$.  The symmetric method is
still much better than the St\"ormer method at most stepsizes, but now there
are spikes in the energy error at a number of stepsizes.  The spikes that
appear at the stepnumbers (the number of steps per orbit) 90, 120, 150,
etc.\ are similar to the spike at 60 steps per orbit: they are instabilities
at which the error grows exponentially with time.  The smaller spikes at
stepnumbers that are multiples of 5 or 6 (54, 55, 65, 66, 70, 72, etc.) are
resonances, and not instabilities, since at these stepsizes the energy error
grows linearly with time, as shown in panel~(b) of Figure~\ref{fig-growth}.

The resonances and instabilities do not occur for the linear test
equation~(\ref{eq-sho}) that has been used in previous discussions of
symmetric multistep methods. The present paper is written to explain their
origin, to see if anything can be done to reduce their severity, and to
decide if they render the methods unsuitable for problems like the long-term
planetary integrations considered by Quinlan and Tremaine (1990).

\section{Symmetric multistep methods}  \label{sec-symmetric}

We start by reviewing symmetric multistep methods and some properties of
them that are needed for explaining the resonances and instabilities. These
properties are described in more detail by Henrici (1962) and Lambert and
Watson (1976).  A survey of multistep methods and other methods for
integrating second-order differential equations is given by Coleman (1993).
Practical techniques for reducing roundoff error in long multistep
integrations are described by Quinlan~(1994).

We consider $k$-step multistep methods of the type~(\ref{eq-genmulti}),
where without loss of generality we assume $\alpha_k=1$ and
$|\alpha_0|+|\beta_0|>0$.  With the multistep method we associate the linear
difference operator
\begin{equation}
  L\left[x(t);h\right]=\sum_{j=0}^{k}\left[\alpha_j x(t+j\cdot h)-
  h^2\beta_j x''(t+j\cdot h)\right].
\end{equation}
If $x(t)$ has continuous derivatives of sufficiently high order then
\begin{equation}
  L\left[x(t);h\right]=C_0x(t)+C_1x'(t)h+\cdots+C_qx^{(q)}(t)h^q+\cdots,
\end{equation}
where
\begin{equation}
  C_q = {1\over q!}(0^q\alpha_0+\cdots+k^q\alpha_k)-
   {1\over (q-2)!}(0^{q-2}\beta_0+\cdots+k^{q-2}\beta_k)
\end{equation}
(the second term on the right is absent if $q<2$).
The order $p$ is the integer for which $C_0= \cdots =C_{p+1} =0$,
$C_{p+2}\neq0$.  A method is said to be consistent if its order is at least
1, i.e., if $C_0 = C_1 = C_2 = 0$.  

To discuss the stability of multistep methods we introduce the polynomials
\begin{eqnarray}
  \rho(z) &=& \alpha_k z^k+\alpha_{k-1} z^{k-1}+\cdots+\alpha_0,\\ \sigma(z)
&=& \beta_k z^k +\beta_{k-1} z^{k-1} +\cdots+\beta_0.
\end{eqnarray}
A method is said to be zero-stable if no root of $\rho(z)$ has modulus
greater than one and if every root of modulus one has multiplicity not
greater than two.  A consistent method has $\rho(1) = \rho'(1) = 0$, so for
zero-stability $\rho(z)$ must have a double root at $z=1$.  This is called
the principal root; the other roots are called spurious.  A method is
convergent if and only if it is consistent and zero-stable (convergence
means essentially that $x_n\to x(t)$ as $h\to0$ with $nh=t$).  The order of
a convergent $k$-step method cannot be higher than $k+2$.

The orbital instability of the St\"ormer-Cowell methods can be explained by
a simple example (Lambert and Watson 1976). Consider
equation~(\ref{eq-sho}), whose general solution is the periodic function
$x(t)=A\cos(\omega t) + B\sin(\omega t)$.  Applying the multistep
method~(\ref{eq-genmulti}) to this equation we obtain the difference
equation
\begin{equation} \label{eq-shodiff}
  \sum_{i=0}^k(\alpha_i+H^2\beta_i)x_{n+i}=0,\qquad H=h\omega,
\end{equation}
whose general solution is
\begin{equation}  \label{eq-shosol}
  x_n=\sum_{j=1}^{k}A_jZ_j^n,
\end{equation}
where the $Z_j$ ($j=$1, 2, \ldots, $k$) are the roots, assumed distinct, of the
polynomial 
\begin{equation} \label{eq-Omega}
  \Omega(Z; H^2)=\rho(Z)+H^2\sigma(Z).
\end{equation}
The roots $Z_j$ of $\Omega$ are perturbations of the roots $z_j$ of $\rho$;
let $Z_1$ and $Z_2$ be the perturbations of the principal roots.  The
problem with the St\"ormer-Cowell methods is that, if $k>2$, the roots $Z_1$
and $Z_2$ do not lie on the unit circle. No matter how small $H^2$ is the
orbit grows or shrinks, and the energy error increases linearly with time.

Lambert and Watson (1976) showed that this orbital instability can be
avoided if the multistep coefficients are chosen to be symmetric, as in
equation~(\ref{eq-sydef}).  A multistep method is said to have an interval
of periodicity $(0,H_0^2)$ if for all $0<H<H_0$ the roots of $\Omega(Z;H^2)$
satisfy
\begin{equation} \label{eq-interval}
  |Z_1|=|Z_2|=1,\qquad |Z_j|\leq1 \quad (j=3,\ldots,k).
\end{equation}
Lambert and Watson proved that a convergent multistep method with a non-zero
interval of periodicity must be a symmetric method and must have an even
order.  For a symmetric method the $\leq$ in~(\ref{eq-interval}) can be
replaced by $=$, and hence inside the interval of periodicity the roots all
lie on the unit circle.  The solution~(\ref{eq-shosol}) of
equation~(\ref{eq-sho}) is then guaranteed to be periodic (or
quasi-periodic).

The requirement that a $k$-step symmetric multistep method have order $k$
(for an explicit method) or $k+2$ (for an implicit method) does not
determine the $\alpha$ and $\beta$ coefficients uniquely when $k>2$, 
because for a symmetric method the equations $C_j=0$ with $j$ odd are not
independent of the equations with $j$ even. There are thus some free
coefficients that can be chosen by the user, although their range is
restricted by the requirement of zero-stability. Lambert and Watson (1976)
gave examples of explicit methods with orders 2, 4, and 6, and implicit
methods with orders 4, 6, and 8.  Quinlan and Tremaine (1990) gave examples
of explicit methods with orders 8, 10, 12, and 14.  
Table~1 lists the coefficients of five explicit methods that will be
discussed in what follows.  The methods SY8, SY10, and SY12 are the 8th-,
10th-, and 12th-order methods of Quinlan and Tremaine; SY8A and SY8B
are 8th-order methods that have not previously been published.  
Table~2 lists the spurious roots of $\rho(z)$ for these five methods.

\begin{table}[ht]
\centering
\begin{tabular}{|r|rr|rr|rr|rr|rr|} 
\hline
 &\multicolumn{2}{c|}{SY8}  & \multicolumn{2}{c|}{SY8A} &
  \multicolumn{2}{c|}{SY8B} & \multicolumn{2}{c|}{SY10} &
  \multicolumn{2}{c|}{SY12}\vphm \\ \hline
 &\multicolumn{2}{c|}{$k=8$} & \multicolumn{2}{c|}{$k=8$} &
  \multicolumn{2}{c|}{$k=8$} & \multicolumn{2}{c|}{$k=10$}&
  \multicolumn{2}{c|}{$k=12$} \vphm \\
 &\multicolumn{2}{c|}{$H_0^2=0.52$} & \multicolumn{2}{c|}{$H_0^2=0.73$} &
  \multicolumn{2}{c|}{$H_0^2=0.10$} & \multicolumn{2}{c|}{$H_0^2=0.17$} &
  \multicolumn{2}{c|}{$H_0^2=0.046$}  \\
 &\multicolumn{2}{c|}{$C_{10}=0.063$} & \multicolumn{2}{c|}{$C_{10}=0.063$} &
  \multicolumn{2}{c|}{$C_{10}=0.059$} & \multicolumn{2}{c|}{$C_{12}=0.058$} &
  \multicolumn{2}{c|}{$C_{14}=0.056$} \\ \hline 
 $i$ &$\alpha_i$&$12096\beta_i$ & $\alpha_i$&$15120\beta_i$ &
 $\alpha_i$&$120960\beta_i$ & $\alpha_i$&$241920\beta_i$
                            & $\alpha_i$&$53222400\beta_i$ \vphm \\ \hline
0&$ 1$ &$  0   $&$ 1$ &$0      $&$ 1  $&$      0 $ &$ 1 $&$       0$ 
                                                   &$ 1 $&$       0$ \vphm\\
1&$-2$ &$ 17671$&$-2$ &$ 22081 $&$ 0  $&$ 192481 $ &$-1 $&$  399187$ 
                                                   &$-2 $&$   90987349$ \\
2&$ 2$ &$-23622$&$ 2$ &$-29418 $&$ 0  $&$   6582 $ &$ 1 $&$ -485156$ 
                                                   &$ 2 $&$ -229596838$ \\
3&$-1$ &$ 61449$&$-2$ &$ 75183 $&$-1/2$&$ 816783 $ &$-1 $&$ 2391436$ 
                                                   &$-1 $&$  812627169$ \\
4&$ 0$ &$-50516$&$ 2$ &$-75212 $&$ -1 $&$-156812 $ &$ 1 $&$-2816732$ 
                                                   &$ 0 $&$-1628539944$ \\ 
5&$  $ &$      $&$  $ &$       $&$    $&$        $ &$-2 $&$ 4651330$ 
                                                   &$ 0 $&$ 2714971338$ \\ 
6&$  $ &$      $&$  $ &$       $&$    $&$        $ &$   $&$$ 
                                                   &$ 0 $&$-3041896548$ \\ 
\hline
\end{tabular}
\caption[Symmetric multistep coefficients]{Coefficients of the symmetric
multistep methods (only half are listed; the others are given by
$\alpha_i=\alpha_{k-i}$, $\beta_i=\beta_{k-i}$).}
\label{tab-coef}
\end{table}

\begin{table}[ht]
\centering
\begin{tabular}{|r|l|c|} \hline
Method & $z=\exp\left(\pm 2\pi i/ n\right)$ & 
$\displaystyle{\max_{j\neq l} {2n_jn_l\over n_j-n_l}}$\\ \hline
SY8    &  $n=$ 2.500, 5.000, 6.000 & 60.00\\
SY8A   &  $n=$ 2.667, 4.000, 8.000 & 16.00\\
SY8B   &  $n=$ 2.278, 3.353, 4.678 & 23.67\\
SY10   &  $n=$ 2.333, 3.500, 6.000, 7.000 & 84.00 \\ 
SY12   &  $n=$ 2.250, 3.000, 4.500, 6.000, 9.000 & 36.00 \\ \hline
\end{tabular}
\caption[Spurious roots for symmetric multistep methods.]{Spurious Roots of
$\rho(z)$ for the methods in Table~\ref{tab-coef}.}
\label{tab-roots}
\end{table}

\section{Origin of the resonances and instabilities} \label{sec-origin}

The origin of the resonances and instabilities will be explained using a
Kepler orbit as an example. The resonances and instabilities occur for all
nonlinear oscillatory problems, not just for the Kepler problem; other
examples will be given later.

\subsection{A simple explanation for the instabilities}

We start with a simple explanation for the instability that occurs when a
circular orbit is integrated with the method SY8 using 60 steps per orbit.
The spurious roots of $\rho(z)$ for the method SY8 are located on the unit
circle at angles $\pm4\pi/5$, $\pm2\pi/5$, and $\pm2\pi/6$ (i.e., at
$\pm144\deg$, $\pm72\deg$, and $\pm60\deg$). At 60 steps per orbit the
spurious roots of $\Omega(Z;\omega^2h^2)$ differ little from those of
$\rho(z)$; the difference will be ignored here.  It is the roots at $2\pi/5$
and $2\pi/6$ (or $-2\pi/5$ and $-2\pi/6$) that cause the trouble.  The root
at $2\pi/5$ allows a 5-step oscillation: in the absence of any forces,
equation~(\ref{eq-genmulti}) admits a solution $x_n=\exp(2\pi in/5)$.
Similarly, the root at $2\pi/6$ allows a 6-step oscillation.  By themselves
the oscillations would be harmless for this problem as long as the start-up
routine is accurate. The trouble arises when the 5- and 6-step oscillations
can resonate.

Consider first the 5-step oscillation.  The perturbations to the $x$ and $y$
coordinates can both oscillate with a 5-step period.  If the $y$ oscillation
is $90\deg$ ahead of the $x$ oscillation, the perturbation goes around in a
counter-clockwise sense with a 5-step period.  Assume that the orbit also
moves in a counter-clockwise sense.  During one orbital period the
perturbation goes around $60/5=12$ times.  Because the perturbation goes
around in the same sense as the orbit, however, the perturbation to the
orbital radius completes only $12-1=11$ oscillations.  Now consider the
6-step oscillation.  This time assume that the $y$ oscillation is $90\deg$
behind the $x$ oscillation, so that the perturbation goes around in a
clockwise sense with a period of 6 steps.  During one orbital period the
perturbation goes around $60/6=10$ times, but the radial perturbation
completes $10+1=11$ oscillations.  This leads to the resonance: the radial
perturbations from both the 5-step and 6-step oscillations can go around 11
times in one orbital period. The perturbation analysis given below shows
that the resonance causes an instability.

The explanation was verified by checking that the instability can be
enhanced by adding noise to the initial conditions with the right frequency
and phase. When noise was added with a 5-step component polarized in a
counter-clockwise sense and a 6-step component polarized in clockwise sense,
the instability became obvious sooner, but when the polarizations were
reversed, so that the 5-step component was clockwise and the 6-step
component counter-clockwise, the instability was delayed.

The explanation predicts that instabilities will occur for other symmetric
multistep methods at stepsizes at which the number of steps per orbit $N$
satisfies
\begin{equation}                                     \label{eq-predict}
 {N\over2\pi/\theta_l}+1= {N\over2\pi/\theta_j}-1,
\end{equation}
where $\exp(i\theta_l)$ and $\exp(i\theta_j)$ are spurious roots of
$\rho(z)$.  The order of the method must be at least six for an instability
of this type to occur, since the polynomial $\rho(z)$ must have at least
four spurious roots.  According to the prediction the method SY10 of Quinlan
and Tremaine should be unstable for circular orbits at 84 steps per orbit,
and the method SY12 should be stable as long as the number of steps per
orbit is at least 36; both predictions were verified, along with similar
predictions for other symmetric multistep methods.

For eccentric Kepler orbits equation~(\ref{eq-predict}) must be generalized
to allow for the different frequency components present in the motion.  The
forces in an eccentric Kepler orbit (with unit major axis) can be expanded
as (Kovalevsky 1967)
\begin{eqnarray}                                          \label{eq-xexpand}
  {x(t)\over r(t)^3} &=& \sum_{q=1}^{\infty}q
          \left[J_{q+1}(qe)-J_{q-1}(qe)\right]\cos(q\omega t),\\ 
                                                          \label{eq-yexpand}
  {y(t)\over r(t)^3} &=& -\sqrt{1-e^2}\sum_{q=1}^{\infty}q
          \left[J_{q+1}(qe)+J_{q-1}(qe)\right]\sin(q\omega t),
\end{eqnarray}
where the $J_q$ are Bessel functions of the first kind.  The forces can be
built up from components with frequencies that are integral multiples of the
fundamental frequency $\omega$. The ``1'' on the left-hand side of
equation~(\ref{eq-predict}) must therefore be allowed to take on the integer
values 1, 2, 3, \ldots, corresponding to the fundamental frequency and the
higher harmonics, and similarly the ``1'' on the right-hand side must be
allowed to take on the same values, independently of the value assumed on
the left-hand side. This leads to the prediction of instabilities at $N=60$,
90, 120, 150, etc. The width of the instability at small eccentricities
(when plotted as in Figure~\ref{fig-sy8}) is found to vary with the
eccentricity as $\sim e^0$ at $N=60$, as $\sim e^1$ at $N=90$, as $\sim e^2$
at $N=120$, and so on, as expected since the different frequency components
have amplitudes that vary as powers of the eccentricity.

\subsection{Perturbation analysis for a circular orbit} \label{sec-perturb}

A perturbation analysis can be used to show that circular orbits are
unstable when the condition~(\ref{eq-predict}) is satisfied.  Consider a
circular orbit of radius $R$ in a two-dimensional axisymmetric potential
$\phi(r)$.  The circular frequency $\omega$ is
\begin{equation} \label{eq-circ}
  \omega(r)^2=\phi'(r)/r.
\end{equation}
Let $x_n$ and $y_n$ be the $x$ and $y$ coordinates at the $n$th time
step. Define $z=x+iy$, $r=|z|$, and write equation~(\ref{eq-genmulti}) as
\begin{equation} \label{eq-multistep}
  \sum_{j=0}^{k}\alpha_j z_{n+j}=h^2\sum_{j=0}^{k}\beta_j F_{n+j},
\end{equation}
where the complex force $F$ is 
\begin{equation} \label{eq-force}
  F=f_x+if_y=-{z\over r}\phi'(r).
\end{equation}
Provided that the stepsize $h$ lies within the interval of periodicity, and
that the radius is assumed to be fixed, so that
equation~(\ref{eq-multistep}) is linear in $z$, the principal roots
$Z_p^{\pm1}$ (in fact all the roots) of~(\ref{eq-multistep}) will lie on the
unit circle, and the numerical solution will have the form
\begin{equation} \label{eq-unperturb}
  z_n=RZ_p^n, \qquad\quad Z_p\simeq e^{i\omega h},
\end{equation}
where $Z_p$ is the principal root of $\Omega(Z;\omega^2h^2)$ corresponding
to the assumed counter-clockwise rotation.  

Now consider a perturbed circular orbit
\begin{equation} \label{eq-perturb}
  z_n=RZ_p^n(1+u_n),
\end{equation}
where $RZ_p^nu_n$ is a small perturbation at time step $n$.  The perturbed
radius is 
\begin{equation} \label{eq-perturbr}
  R_n=(z_nz_n^*)^{1/2}\simeq R[1+(u_n+u_n^*)/2],
\end{equation}
where the $^*$ denotes complex conjugation.  Substituting~(\ref{eq-perturb})
into equation~(\ref{eq-multistep}) and linearizing the resulting equation we
find
\begin{equation} \label{eq-linear}
  \sum_{j=0}^{k}\alpha_j Z_p^{j}u_{n+j}=-h^2\sum_{j=0}^{k}\beta_jZ_p^{j}
    \left(\omega_1u_{n+j}-\omega_2u_{n+j}^*\right),
\end{equation}
where
\begin{eqnarray} \label{eq-w1def}
  \omega_1 &=& {1\over2}\left[{1\over r}\phi'(r)+\phi''(r)\right]\ =\ 
    {1\over2}(\kappa^2-2\omega^2),\\ \label{eq-w2def}
  \omega_2 &=& {1\over2}\left[{1\over r}\phi'(r)-\phi''(r)\right]\ =\ 
    {1\over2}(4\omega^2-\kappa^2),
\end{eqnarray}
and where $\kappa$ is the epicyclic frequency, given 
by (Binney and Tremaine 1987) 
\begin{equation} \label{eq-kappa}
  \kappa^2(r)=r {d\omega^2\over dr}+4\omega^2=\phi''(r)+{3\over r}\phi'(r).
\end{equation}
A trial perturbation of the form
\begin{equation} \label{eq-dispuc}
  u_n=A_1S^n+A_2S^{*n}
\end{equation}
leads to a solution for $S$ that is independent of the step number $n$ if
the following two equations are satisfied: 
\begin{eqnarray}
  A_1\sum_{j=0}^{k}\left(\alpha_jZ_p^jS^j+\omega_1h^2\beta_jZ_p^jS^j\right) &=&
  A_2^*\sum_{j=0}^{k}\omega_2h^2\beta_jZ_p^jS^j;\\
  A_2\sum_{j=0}^{k}\left(\alpha_jZ_p^jS^{*j}+\omega_1h^2\beta_jZ_p^jS^{*j}
    \right) &=&
  A_1^*\sum_{j=0}^{k}\omega_2h^2\beta_jZ_p^jS^{*j}.
\end{eqnarray}
Taking the complex conjugate of the second equation, we get two equations for
the two unknowns $A_1$ and $A_2^*$. The determinant of the system must vanish,
which gives (using $H=\omega h$)
\begin{equation} \label{eq-dispo}
  D(S)=\Omega(SZ_p;H^2)\Omega(S/Z_p;H^2) - 
  \omega_2h^2
  \left[\Omega(SZ_p;H^2)\sigma(S/Z_p) +
  \Omega(S/Z_p;H^2)\sigma(SZ_p) \right]=0.
\end{equation}

$D(S)$ is a symmetric polynomial of degree $2k$ in $S$, with real
coefficients.  There are several things to note about the roots of this
polynomial. If $S$ is a root then so are $S^*$, $1/S$, and $1/S^*$.  $S=1$
is always a root, since $\Omega(Z_p;H^2) = \Omega(1/Z_p;H^2)=0$. The root
$S=1$ corresponds to unperturbed motion, because for this root $A_1=-A_2^*$
and hence the linearized perturbation to the radius is zero.  If $h=0$ then
$D(S)=\rho(S)^2$, whose roots are the same as those of $\rho(z)$, with each
root $S_i$ having twice the multiplicity of the corresponding root $z_i$ of
$\rho(z)$.  Thus when $h=0$ the spurious roots of $\rho(z)$ (assumed to be
distinct) are double roots of $D(S)$, and $S=1$ is a root of multiplicity
four.  If $h$ is small but nonzero the roots that were double roots of
$D(S)$ when $h=0$ split, and are approximately $Z_jZ_p^{\pm1}$
$(j=3,\ldots,k)$, where the $Z_j$ are the spurious roots of $\Omega(Z;H^2)$.
The root $S=1$ is a double root when $h\neq0$, and two new roots appear at
approximately $Z_p^{\pm\kappa/\omega}$, corresponding to the usual epicyclic
oscillations with frequency $\omega\pm\kappa$.

As $h$ increases away from $0$ the roots of $D(S)$ move around the unit circle
as just described.  Problems arise, however, near
stepsizes where $Z_p^2Z_l/Z_j=1$ for some spurious roots $Z_l$ and $Z_j$.
When this happens the approximate solutions for two of the roots coincide,
$Z_lZ_p= Z_j/Z_p$, and a more careful analysis is needed. The troublesome
stepsizes can be estimated by replacing $Z_p$ by $\exp(i\omega h)$ and $Z_l$ and
$Z_j$ by the corresponding roots of $\rho(z)$, which we write as
$z_l=\exp(i\theta_l)$ and $z_j=\exp(i\theta_j)$. The troublesome stepsizes
then occur when
\begin{equation}                                       \label{eq-trouble}
  \theta_j-\theta_l=2\omega h=4\pi/N,
\end{equation}
where $N$ is the number of steps per orbit.  This is the instability
criterion~(\ref{eq-predict}) that was given earlier.

\begin{figure}[ht]
\centerline{\psfig{figure=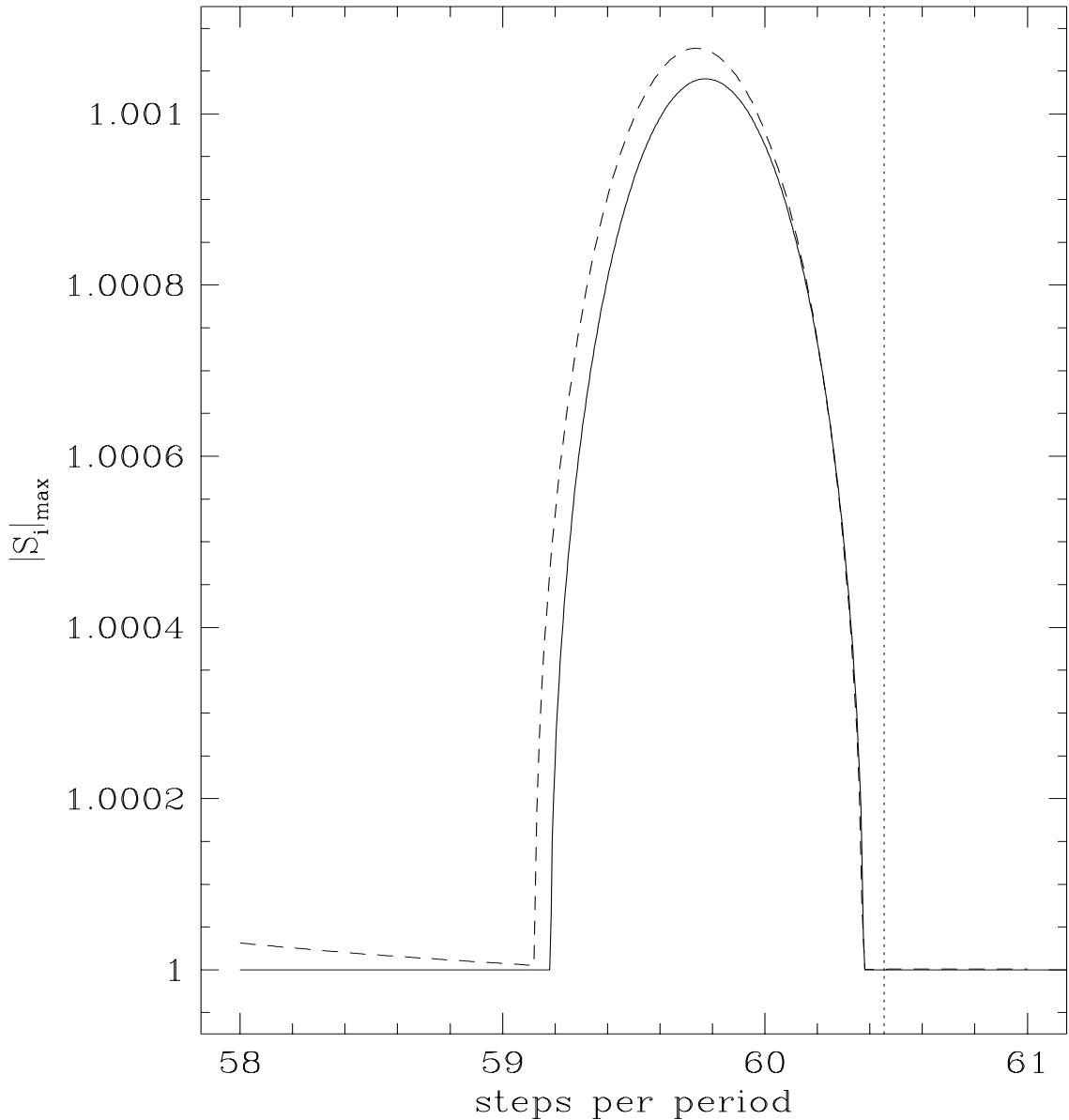,width=\textwidth,height=0.85\textheight}}
\vskip-40pt
\caption[Roots of $D(S)$]
{Maximum value of $|S_i|$ ($i=1,\ldots,16$) for the roots of $D(S)$ using the
method SY8 with a Kepler potential $\phi(r)=-1/r$.
The solid line is the exact result from a numerical calculation of
the roots; the long-dashed line is the approximate solution described in the
text.  The short-dashed line marks the value $N\simeq60.455$ at which
$Z_p^2Z_l/Z_j=1$ for two of the spurious roots of $\Omega$.}
\label{fig-sroot}
\end{figure}

Figure~\ref{fig-sroot} shows the maximum magnitude of the roots of $D(S)$
for the method SY8 used near 60 steps per orbit with a Kepler potential
$\phi(r)=-1/r$.
The solid line is the exact solution found by numerical calculation of the
roots.  The dashed line is an approximate solution found by simplifying
$D(S)$ by writing   
\begin{equation}
 \Omega(SZ_p^{\pm1};H^2)\simeq\prod_{m=1}^{8}(SZ_p^{\pm1}-Z_{m,r}),
\end{equation}
where the $Z_{m,r}$ are the roots of $\Omega$ at the resonant stepsize
($N\simeq60.455$), and by replacing $SZ_p$ by $Z_{l,r}$ and $S/Z_p$ by
$Z_{j,r}$ everywhere except in the terms $(SZ_p-Z_{l,r})$ and
$(S/Z_p-Z_{j,r})$; the same replacements are made in the $\sigma$
functions. We thus obtain from $D(S)=0$ a quadratic equation in $S$, whose
roots are easily found. The exact and approximate solutions both show that
there are roots of $D(S)$ that lie off the unit circle when $N$ is in the
range 59.2--60.4.  The amount by which the roots move off the unit circle
depends on the value of $\kappa^2-4\omega^2$.  For a harmonic oscillator
potential $\phi(r)\sim r^2$ the roots of $D(S)$ do not move off the unit
circle, since $\kappa^2-4\omega^2$=0, and the instability does not occur.

\subsection{Location of the resonances}

The resonances in Figure~\ref{fig-sy8} are easier to predict
than the instabilities. Since the force acting on an eccentric Kepler orbit
is a superposition of components with frequencies that are integral
multiples of the fundamental frequency $\omega$, the resonances occur when the
principal root $Z_p\simeq\exp(i\omega h)$ raised to some integral power $q$
coincides with one of the spurious roots $Z_j$, i.e., when
\begin{equation}                                          \label{eq-epredict}
   N \simeq {2\pi q\over \theta_j}, \qquad\qquad q=1,2,3,\ldots,
\end{equation}
where $z_j=\exp(i\theta_j)$ is a spurious root of $\rho(z)$. This correctly
predicts the resonances for the method SY8 at stepnumbers that are multiples
of 5 and 6 (and also 2.5, although these resonances are usually too weak to
be seen). The amplitude and growth rate of the resonance decrease as $q$
increases, because for small eccentricities the Bessel functions in
equations~(\ref{eq-xexpand}) and (\ref{eq-yexpand}) decrease rapidly with
$q$.

\section{Further examples} \label{sec-examples}

Three more examples will be given to show how the resonance and instability
locations depend on the multistep method and on the equation being integrated.

\subsection{Kepler orbits with a 12th-order symmetric method} \label{sec-sy12}

\begin{figure}[ht]
\centerline{\psfig{figure=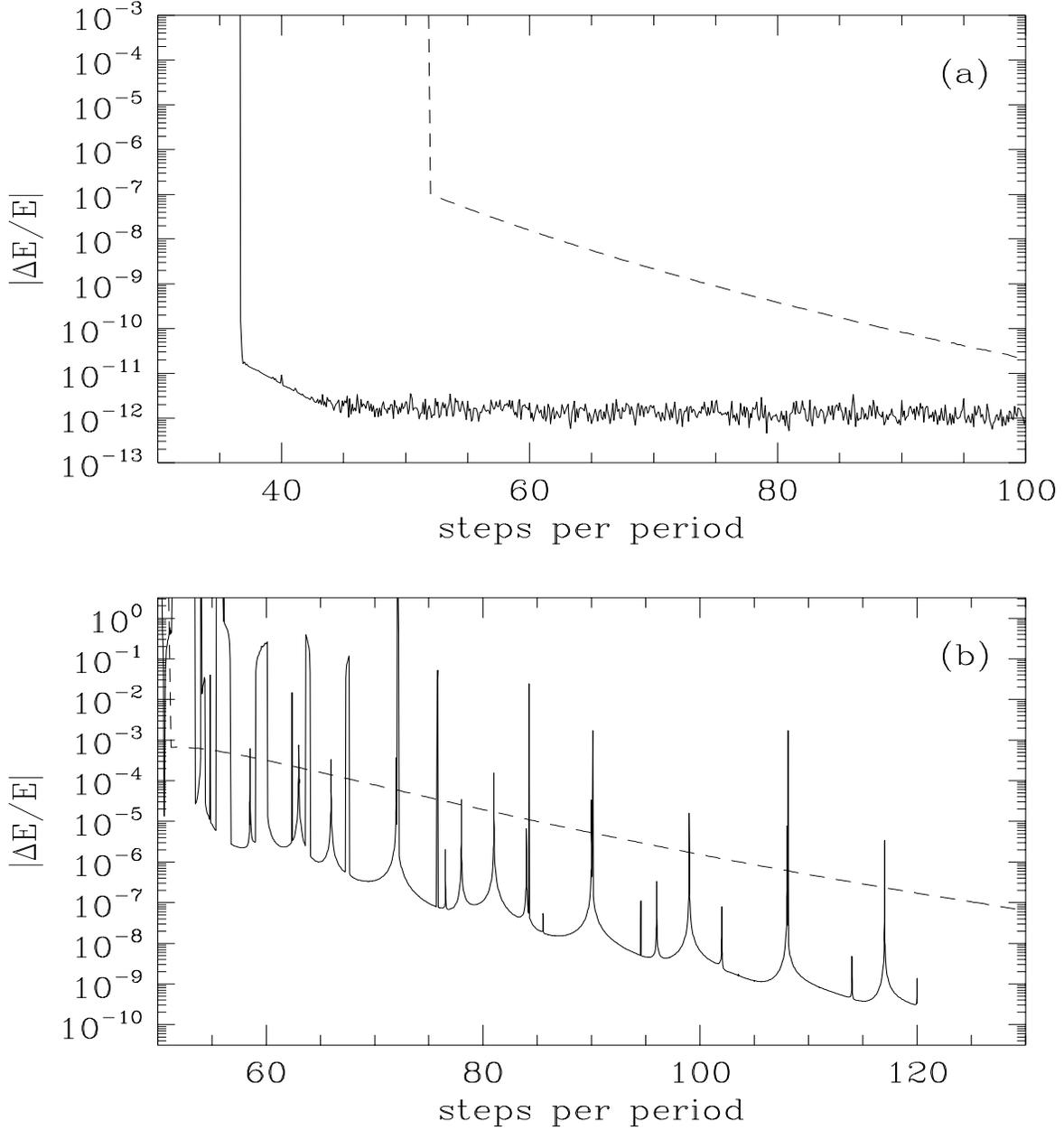,width=\textwidth,height=0.85\textheight}}
\vskip-30pt
\caption[SY12] {Maximum fractional energy error during integrations of a
Kepler orbit for 25000 periods using the method SY12 (solid lines) and the
12th-order St\"ormer method (dashed lines) with about 1600 different
stepsizes, plotted as a function of the number of steps per orbit: (a)
circular orbit; (b) eccentric orbit ($e=0.2$).}
\label{fig-sy12}
\end{figure}

The 12th-order symmetric method SY12 is expected to be better than the
8th-order method SY8 for circular Kepler orbits, since the 12th-order method
is free of instabilities for these orbits as long as the integration takes
at least 36 steps per orbit. This is confirmed in panel~(a) of
Figure~\ref{fig-sy12} (the energy error of the symmetric method is caused by
roundoff error at most stepsizes in this panel).
For eccentric orbits, however, the 12th-order method can be as troublesome
as the 8th-order method, if not more so, because the extra spurious roots
allow more opportunities for resonances and instabilities to occur, and
because the spurious root at $2\pi/9$ on the unit circle leads to a
resonance that can be excited even at low eccentricities. Panel~(b) of
Figure~\ref{fig-sy12} shows resonances at stepnumbers that are multiples of
4.5, 6, and 9, and instabilities at stepnumbers that are multiples of 36,
although these are not the only instabilities; there are others in the
figure, and more would be present if the stepsizes had been sampled more
finely.
	
\subsection{Orbits in a logarithmic potential}                 \label{sec-eccL}

\begin{figure}[ht]
\centerline{\psfig{figure=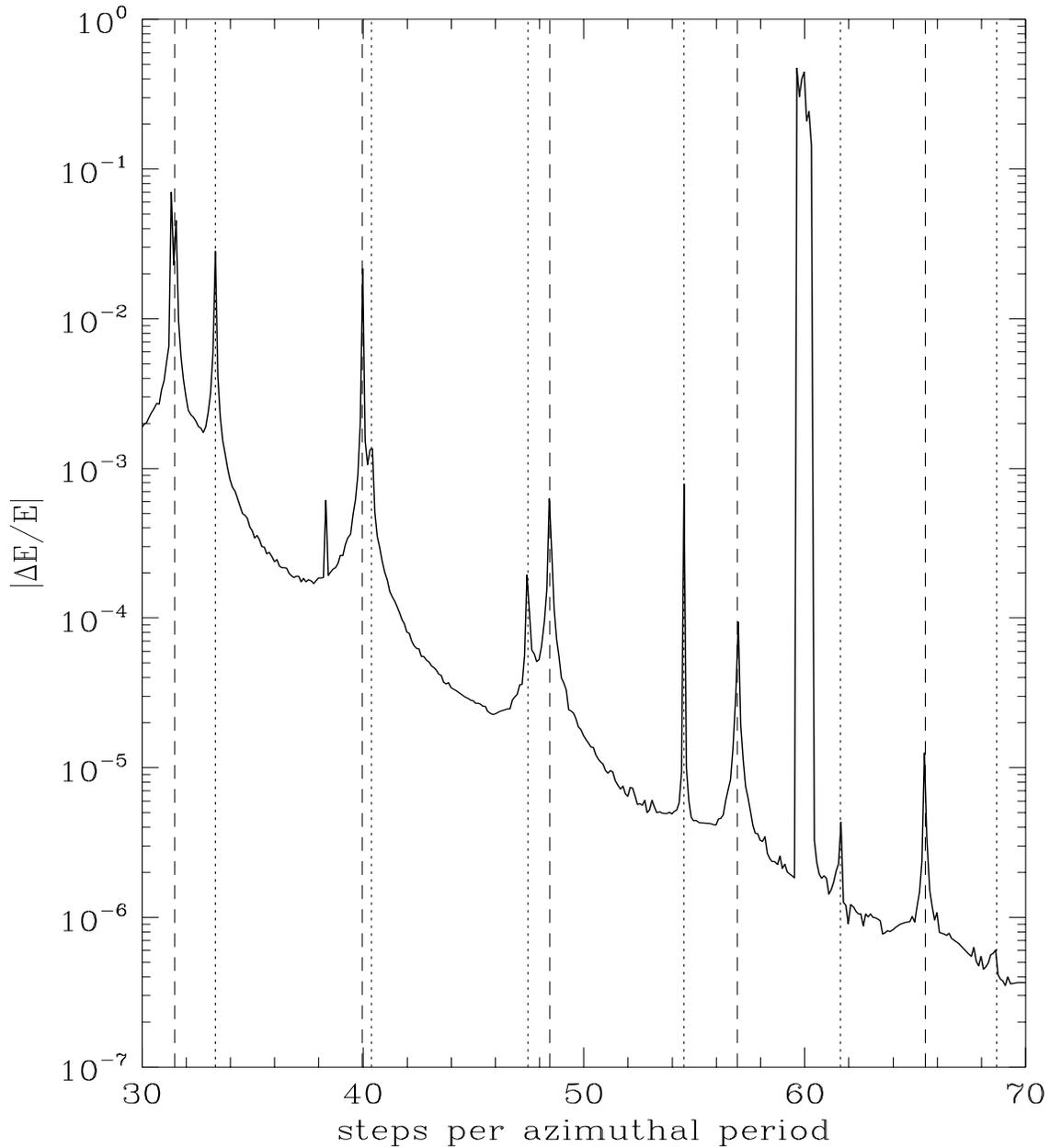,width=\textwidth,height=0.85\textheight}}
\vskip-30pt
\caption[logr]
{Maximum fractional energy error during integrations of an eccentric orbit
in a logarithmic potential for 10000 periods with the method SY8,
using 500 equally-spaced stepsizes (see text).}
\label{fig-logr}
\end{figure}

The Kepler potential is special in that all bound orbits are closed and have
$\omega=\kappa$, i.e., the azimuthal period $T_a=2\pi/\omega$ is equal to
the radial period $T_r=2\pi/\kappa$ (equation~(\ref{eq-kappa}) gives $\kappa$
for a circular orbit).  For most potentials $T_a\neq T_r$, and the analysis
given previously must be modified.  Consider an eccentric orbit in a
logarithmic potential $\phi(r)=\log(r)$, for which (in the limit of a
circular orbit) $\kappa=\sqrt{2}\omega$. A Fourier analysis of the motion
contains frequencies $\omega+q\kappa$ (where $q$ is an integer), and hence
the prediction~(\ref{eq-epredict}) for the resonant stepsizes must be
modified to read
\begin{equation}                                         \label{eq-epredictg}
  N\simeq{2\pi\over\theta_j}\left(1+{q\kappa\over\omega}\right)=
  {2\pi\over\theta_j}\left(1+{qT_a\over T_r}\right),\qquad q=0,1,2,\ldots,
\end{equation}
where $N=T_a/h$ is the number of steps per azimuthal period.  This
prediction is verified for the method SY8 in Figure~\ref{fig-logr}, which
shows results from integrating an orbit with initial conditions $x_0=1$,
$y_0=0$, ${x'}_0=0$, ${y'}_0=1.1$, for which $T_a/T_r=1.41536$, close to the
value $\sqrt{2}$ for a circular orbit (the stepsizes were not sampled as
finely in this figure as in the other figures).
There is an instability at 60 steps per azimuthal period, just as with the
Kepler potential, and resonances at the $N$ values predicted by
equation~(\ref{eq-epredictg}), taking $\theta_j$ to be $2\pi/5$ (the short
dashed lines in Figure~\ref{fig-logr}) or $2\pi/6$ (the long dashed lines).

\subsection{The one-dimensional hard spring}             \label{sec-hardspring}
\begin{figure}[ht]
\centerline{\psfig{figure=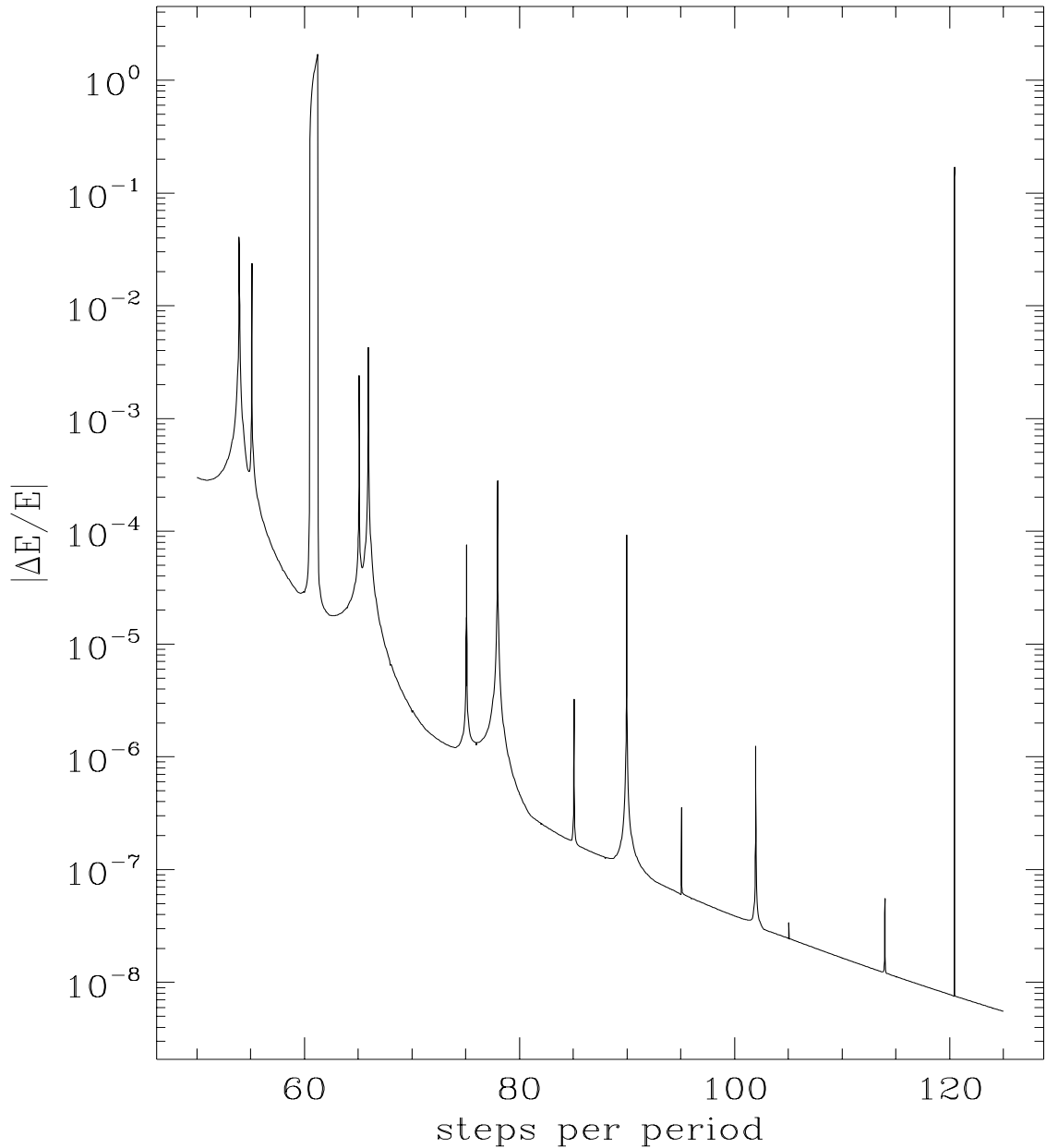,width=\textwidth,height=0.85\textheight}}
\vskip-30pt
\caption[Hard-spring results.]{Maximum energy error measured during
integrations of the hard-spring equation~(\ref{eq-hardspring}) for
50000 periods starting from the initial conditions $x_0$=0 and
$x'_0=1/\sqrt{2}$, using the method SY8 with 1600 different stepsizes.}
\label{fig-hardspring}
\end{figure}

As a final example we consider the one-dimensional hard-spring equation
\begin{equation}                                          \label{eq-hardspring}
  x''(t)=-x(t)^3.
\end{equation}
While this is not an orbital equation, the motion is similar to an eccentric
orbit in that it is periodic and contains frequencies that are integral
multiples of the fundamental frequency, although in this case only the odd
multiples are present. Figure~\ref{fig-hardspring} shows the errors obtained
by integrating this equation with the method SY8.
As expected, there are resonances at stepnumbers that are odd multiples of 5
or 6, such as 65, 66, 75, 78, etc.; the even multiples are missing because
of the missing frequencies in the motion. There are instabilities at 60
steps per orbit and at integer multiples of this number: 120, 180, etc.; the
instabilities at 90, 150, etc., are missing, again because of the missing
frequencies.

\section{A search for better symmetric methods} \label{sec-better}

A question that naturally arises is whether Quinlan and Tremaine (1990)
could have chosen better multistep coefficients to reduce the resonance and
instability problems. The answer is yes for the 8th- and 10th-order methods,
but probably no for the 12th-order method.

There are three properties that we would like a symmetric multistep method
to have: 
\begin{enumerate}
\item The method should have a large interval of periodicity. \label{item-interval}
\item To avoid instabilities the spurious roots of $\rho(z)$ should be well 
      spread out on the unit circle. \label{item-instab}
\item To avoid resonances the spurious roots of $\rho(z)$ should be far 
      from $z=1$. \label{item-res}
\end{enumerate}
Fukushima~(1998) has searched for multistep methods with large intervals of
periodicity. It is properties \ref{item-instab} and \ref{item-res} that
concern us here.  Unfortunately these properties are not compatible: the
spurious roots cannot be well spread out on the unit circle and at the same
time be far from $z=1$.  A compromise must be made, which becomes more
difficult the more spurious roots there are, i.e., the higher the order of
the method.

A systematic search was made through symmetric multistep methods with
$\alpha$ coefficients drawn from the set \{0, $\pm1/8$,
$\pm2/8$,\ldots,$\pm7/8$,$\pm1$,$\pm2$,$\pm4$\} (with $\alpha_k=1$). This
set was chosen because the integration method suffers less from roundoff
error if the $\alpha$'s are integral powers of 2; it is unlikely that much
better methods would have been found by letting the $\alpha$'s take on
arbitrary values. For each set of coefficients the roots of $\rho(z)$ were
computed and checked to see if and where they lie on the unit circle.  The
most promising methods (based on the three properties given above) were
compared in integrations of eccentric Kepler orbits.

\begin{figure}[ht]
\centerline{\psfig{figure=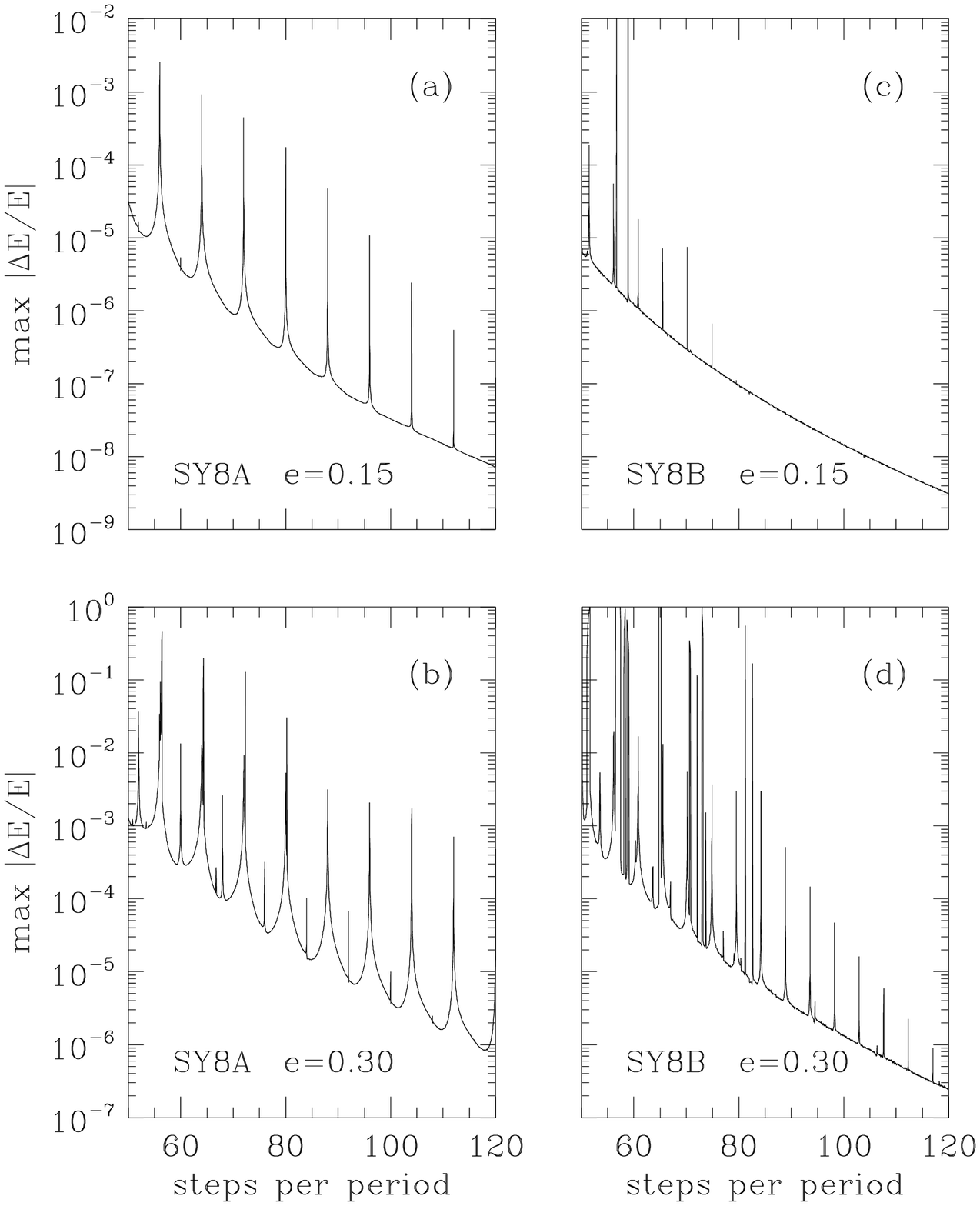,width=\textwidth,height=0.85\textheight}}
\vskip-30pt
\caption[New 8th-order methods.]{Error versus stepsize during
integrations of Kepler orbits for 25000 periods with the 8th-order
symmetric methods SY8A and SY8B, using 1500 different stepsizes.}
\label{fig-sy8ab}
\end{figure}

Consider first the 8th-order methods. For any even integration order $k$,
the choice 1, $-2$, $+2$, $-2$, $+2$, \ldots, for the $\alpha$ coefficients
spreads the spurious roots on the unit circle as evenly as possible, and
tends to minimize the instabilities. That choice was made for the method
SY8A listed in Table~\ref{tab-coef}, which has the largest interval of
periodicity of the 8th-order methods that were tested.  This method is much
more stable than the method SY8; the integration results in
Figure~\ref{fig-sy8ab} show no signs of instabilities even for an orbit with
an eccentricity $e=0.3$.
The drawback of the method SY8A, however, is the spurious root at $2\pi/8$,
or $45^\circ$, which leads to resonances (at stepnumbers $N$ that are
multiples of 8) that are stronger than for SY8.  The method SY8A is
certainly an improvement over SY8, but the resonances are annoying.

The resonances can be reduced by picking a method whose spurious roots are
farther from $z=1$.  An example is the method SY8B, whose spurious root
closest to $z=1$ is at $76.96^\circ$. (It is easy to find methods whose
resonances are weaker than those of SY8B, but these methods have much poorer
stability properties.)  Figure~\ref{fig-sy8ab} shows that the resonances are
much weaker with this method than with SY8A and SY8: they fall off faster as
the number of steps per orbit is increased, being almost unnoticeable when
the $e=0.15$ orbit is integrated with more than 75 steps per orbit, and the
$e=0.30$ orbit with more than 120 steps per orbit.  But the price we pay for
this gain is a decrease in stability, as revealed by the spikes in the error
plot, especially at the higher eccentricity, $e=0.3$.  The instabilities can
be avoided if the stepsize is chosen small enough, however, and are not as
bad as with the method SY8. The method SY8B is the most promising of the
8th-order methods that were tested.

The search for better 10th- and 12th-order symmetric methods was not as
successful.  The method SY10 of Quinlan and Tremaine (1990) is unstable at
84 steps per orbit, even for a circular orbit. The method SY10A (like SY8A,
with the spurious roots spread out evenly on the unit circle) is much more
stable, but has an annoying resonance at stepnumbers $N$ that are multiples
of 10. None of the other 10th-order methods tested was much better than
SY10A.  For the 12th-order methods none was found to be better than SY12.
The method SY12A (again, with the spurious roots spread out evenly on the
unit circle) has an annoying resonance at stepnumbers that are multiples of
12, and its stability properties are no better than those of SY12.

Fukushima~(1999) has tested some implicit symmetric methods to see if they
are affected by the resonance and instability problems as badly as the
explicit methods. The disadvantage of implicit symmetric methods is that the
corrector step must be applied at least twice for the benefits of the
symmetry to be realized, so that for the same stepsize an implicit method
requires at least three times the number of force evaluations required by an
explicit method. Unless implicit methods can be found with much weaker
resonances and instabilities than the explicit methods, this extra work
would be better spent using an explicit method with a smaller stepsize.

\section{The use of symmetric methods for integrating planetary orbits} 
\label{sec-planets}

The examples considered so far have been chosen for their simplicity and
pedagogical value.  We now test the method SY12 on a more complex example,
typical of those encountered in research problems, to get a better idea of
how the method compares with a high-order St\"ormer method and how its
effectiveness is reduced by the resonances and instabilities.  The example
is the long-term integration of a planetary system, the problem that was the
motivation for the high-order symmetric multistep methods of Quinlan and
Tremaine (1990), and a research problem on which the method SY12 has been
used (the 3~Myr integration of all nine planets by Quinn, Tremaine, and
Duncan 1991).  We consider an idealized solar system containing only two
planets, Jupiter and Saturn, as this is sufficiently complex to suggest how
the method will work for a system with more than two planets.

\begin{figure}[ht]
\centerline{\psfig{figure=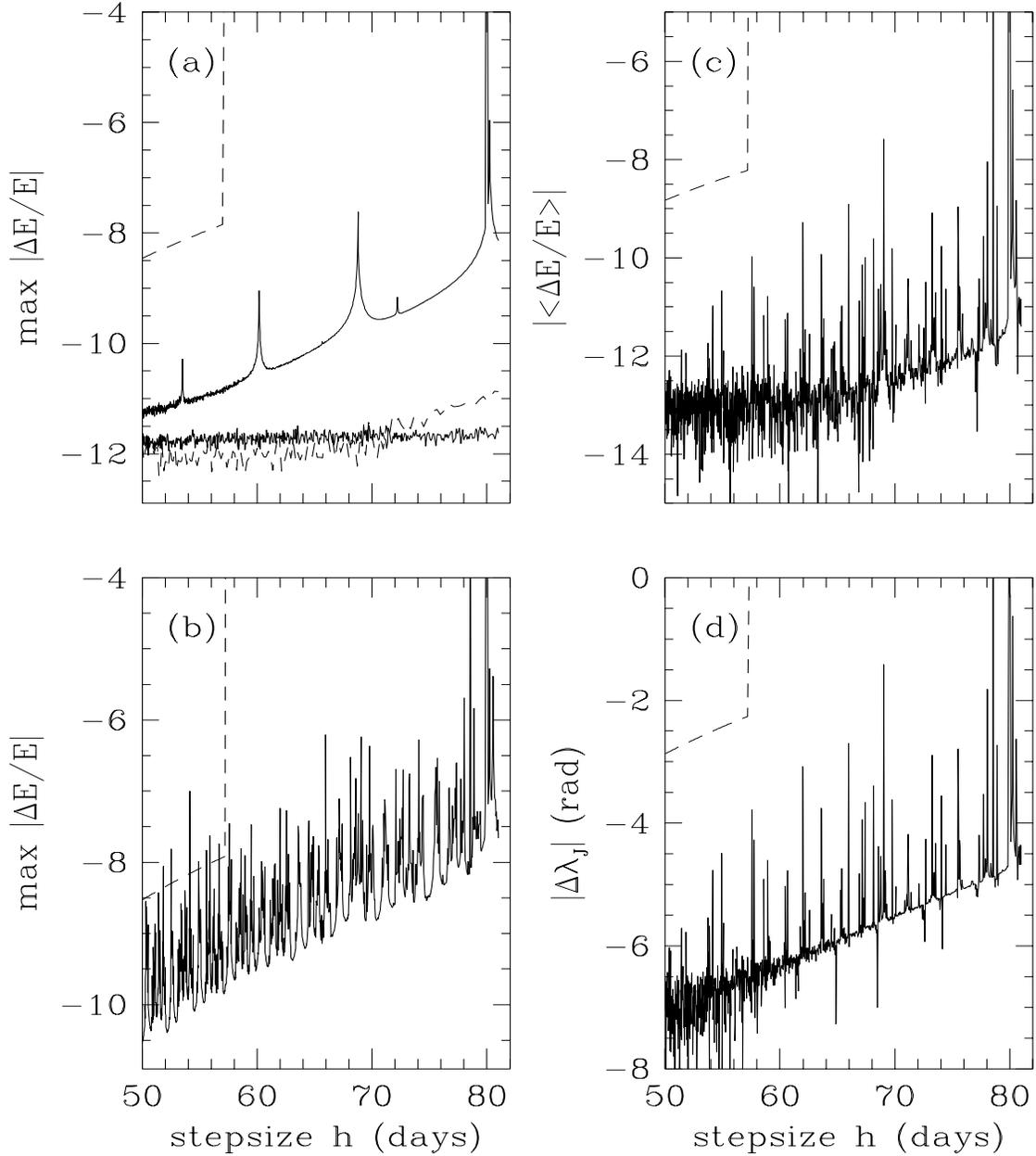,width=\textwidth,height=0.85\textheight}}
\vskip-30pt
\caption[Jupiter-Saturn results.]{Error versus stepsize for 1~Myr
integrations of the Jupiter-Saturn system using the 12th-order symmetric
method SY12 (solid lines) and 13th-order St\"ormer method (dashed lines).
The ordinate gives the common logarithm of the error.  Panel (a) shows the
maximum energy error in the orbits of Jupiter (upper lines) and Saturn
(lower lines) when the planets are integrated separately, with no
interaction between them.  The other three panels include the gravitational
attraction between Jupiter and Saturn, and show the maximum (b) and average
(c) energy errors during the integration, and (d) the longitude error for
Jupiter at the end of the integration.}
\label{fig-js}
\end{figure}

The initial conditions for Jupiter and Saturn are taken from Standish
(1990).  The initial major axes are $a_J\simeq5.203\AU$ and
$a_S\simeq19.280\AU$, giving orbital periods $P_J\simeq4332.8$ and
$P_S\simeq30905$ (in days, the unit of time in this discussion).  The
initial eccentricities are $e_J\simeq0.048$ and $e_S\simeq0.051$.  The
orbits were integrated for 1~Myr using 1000 different stepsizes $h$ spaced
equally in $1/h$ between $h=50$ and $h=81$, corresponding to approximately
86.6 and 53.5 steps per orbit for Jupiter.  To speed the calculations the
energy error was measured every 5th integration step, rather than every
step. Jupiter's longitude at the end of the integration was compared with an
accurate value determined by an integration with a much smaller stepsize
($h=10$). The integration errors are plotted versus stepsize in
Figure~\ref{fig-js}.

The two planets were first integrated separately, with no gravitational
interaction between them.  The maximum energy errors are shown in panel~(a).
The errors for Saturn in this panel are caused by roundoff error and are not
interesting.  The St\"ormer method is unstable if it is used for Jupiter's
orbit with a stepsize larger than $h\simeq57$.  The symmetric method is
stable for Jupiter's orbit with stepsizes as large as $h\simeq80$, although
the effects of resonances are seen when the number of steps per orbit is a
multiple of 9 (near the $h$ values 53.5, 60.2, and 68.8, corresponding to
81, 72, and 63 steps per orbit), and also, to a lesser extent, a multiple of
6 (near $h=72.2$, or 60 steps per orbit).  The spike in the error near
$h=80$ (54 steps per orbit) is an instability, not a resonance, as there is
a sudden growth in the error to a large value early in the integration.
Away from the resonances the maximum energy error is more than 100 times
smaller with the symmetric method than with the St\"ormer method.

The planets were then integrated together including their gravitational
interaction, in the hope that it might detune the resonances and remove the
spikes from the error plot.  This is not what happened.  Panel~(b) shows a
profusion of new resonances resulting from the high-frequency perturbation
terms in the Jupiter-Saturn interaction.  There is still the instability
near $h=80$ that was present for Jupiter alone, plus a narrow instability
near $h=78.55$ that was not present for Jupiter alone, but it is hard to
identify the resonances at 60, 63, 72, and 81 steps per Jupiter orbit, as
they are lost in the other resonances.

With the symmetric method the maximum energy error away from the resonance
peaks is noticeably larger for the Jupiter-Saturn system than for the
single-planet (Jupiter) system, whereas for the St\"ormer method the errors
for the two systems are comparable.  At first this suggests that the
symmetric method is not much better than the St\"ormer method for the two
planet system. But panels~(c) and (d) suggest the opposite conclusion.  With
the symmetric method the average energy error is several orders of magnitude
smaller than the maximum error, whereas with the St\"ormer method the
average and maximum errors are comparable. The error in Jupiter's longitude
at the end of the integration is much smaller with the symmetric method than
with the St\"ormer method.  Even at the resonance peaks the longitude error
is more than an order of magnitude smaller with the symmetric method than
with the St\"ormer method, and away from the resonances the difference is
more than three orders of magnitude.  The narrowness of the resonances is
difficult to appreciate from the figure.  Out of the 310 stepsizes that were
tested in the range $h=60$--70, only 16 (or 8) resulted in longitude errors
larger than $10^{-5}$ (or $10^{-4}$).

These results show that the maximum energy error can be an unduly strict
criterion to use for evaluating the performance of a symmetric method.  In a
planetary integration the position errors are determined by the average
energy error, which for the symmetric methods is much smaller than the
maximum error.  The maximum error is important for calculations of
instantaneous orbital elements that depend on the velocity, such as the
major axis, eccentricity, and inclination.  In the 3~Myr integration of the
solar system by Quinn et al.~(1991) these elements were digitally filtered
to remove oscillations with periods smaller than $500\,$yr; the filtering
probably removed any spurious oscillations caused by the symmetric method,
and the errors in the output elements were probably closer to the average
errors than to the maximum errors. The comparison that Quinn et al.\ made
of their results with the results of a shorter but more accurate St\"ormer
integration (with a smaller stepsize) showed satisfactory agreement.

Despite the resonances and instabilities, then, symmetric methods can
still be a better choice than St\"ormer methods for long integrations
of planetary orbits provided that the user is aware of the
dangers.

\section*{Acknowledgement}

Scott Tremaine and I learned of the resonance and instability problems from
Alar Toomre, who discovered them through numerical experiments soon after
our 1990 paper was published.  Alar's detailed comments on an early draft of
this paper (April 1991) improved my understanding and presentation of the
problem; his further experiments and enthusiastic help since then have been
greatly appreciated.  Scott suggested the perturbation analysis given in
Section~\ref{sec-perturb}, and helped in a number of other ways.  This
research was funded by the Natural Sciences and Engineering Research Council
of Canada (NSERC). Most of the research was completed while I was working at
CITA; I thank Dick Bond for letting me use an office there to finish the paper.


\end{document}